# CREATING AND IMPLEMENTING A SMART SPEAKER

[1]sanskar Jethi, [2]avinash Kumar Choudhary, [3]yash Gupta, [4]abhishek Chaudhary

[1,2,3,4]Department of Electrical Engineering, Delhi Technological University, Bawana Road, Delhi, India.
[1]sansyrox@gmail.com, [2]avinash.ee61@gmail.com, [3]yash04g@gmail.com, [4]abhishek@dtu.ac.in

**Abstract:** We have seen significant advancements in Artificial Intelligence and Machine Learning in the 21st century. It has enabled a new technology where we can have a human-like conversation with the machines. The most significant use of this speech recognition and contextual understanding technology exists in the form of a Smart Speaker. We have a wide variety of Smart Speaker products available to us. This paper aims to decode its creation and explain the technology that makes these Speakers, "Smart."

**1.     Introduction**

The progressions in semiconductor innovation have decreased measurements and cost while improving the exhibition and limit of chipsets. Moreover, advancement in the AI structures and libraries carries prospects to oblige more AI at the asset compelled edge of purchaser IoT devices. These progressions made it feasible for buyer electronic devices, for example, smart speakers to be made in little structure factors, yet equipped for running unique calculations to catch measures and comprehend voice orders. Such devices are incredibly affecting our everyday life. Sensors are these days an integral part of our environment which provides nonstop information streams to construct shrewd applications. A model could be a smart home scenario with numerous interconnected devices. In a smart home scenario, multiple smart devices ( for example, smart security cameras, video doorbells, smart attachments, smart carbon monoxide screens, smart entryway bolts, and alarms, and so on) are interlinked and work in a joint effort with one another to serve a shared objective. Smart speakers are among such intelligent devices that are by and large broadly received by regular clients and turning into a fundamental piece of smart homes.

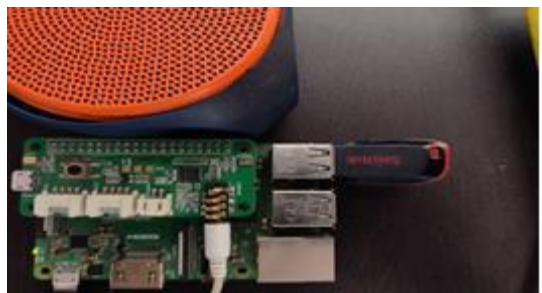

Fig 1. Kasper Smart Speaker

To decipher these technologies' internal working and gain a better insight into these devices, we have tried to create our smart speaker. We call it Kasper.AI.

We are using a raspberry pi as our primary computing unit with a reSpeaker 2 mic array and a speaker module for audio i/o. All the computational and file processing work will be done on-board on the raspberry pi. All components used are targeted to be cross-platform and are not too generic for a board or hardware set. We have implemented the Smart Assistant client through a modified finite state machine architecture. The state machine will be started on hot word detection and all the states will be covered as the subsequent events occur. The brain/logic of the Smart Speaker will be aided through Recurrent Neural Networks, REST API architecture, and crowdsourced datasets to generate and serve intelligent responses.

Inbuilt within our smart speakers can understand voice-based commands and control complex integrated systems of a smart home. Users start interacting with a regular smart speaker by waking up the Kasper voice assistant by calling out the "Kasper" wake-word, followed by regular dialogues-based interaction. Currently, developers have implemented AI algorithms that focus on improving the performance of conversational AI systems. This article describes the design and development of state-of-the art, Linux-based modern smart speaker prototype. The smart speaker discussed in this paper is constructed using off-the-shelf hardware components (Raspberry Pi, ReSpeaker v2, Raspberry Pi camera, regular speaker). In this work, to provide a seamless, full-duplex interaction, a microphone array with an on-board chip hosting DSP-based speech algorithms was selected and used to capture, process, and provide a noise suppressed voice feed. As a result, our proof of concept prototype demonstrates a rich user experience to interact with smart speakers by improving voice interaction with the device. Recent relevant work is done by authors from. We have used the same ReSpeaker v2 microphone-array to provide advanced voice interaction capability for their microphone array and voice algorithm-based smart hearing aid prototype.

## 2. Methodologies

### A. Hardware Components
- Raspberry Pi
- ReSpeaker 2-mics Hat / USB mic / USB sound card
- SD card
- speaker
- 3.5mm Aux cable/ JST PH2.0 connector

A variety of alternatives were chosen before finalising these components.

**Alternatives Considered:**

The Raspberry Pi has a couple of contenders, albeit the establishment urges individuals to clone its thought, so rival probably won't be the correct word. They incorporate BeagleBoard and PandaBoard (which are both the organizations' names and their essential gadgets). Both are charitable associations yet with somewhat unexpected objectives in comparison to the Raspberry Pi Foundation. BeagleBoard is intended for grown-up equipment hobbyists, and PandaBoard expects to make a versatile programming stage accessible at a sensible cost.

Like Raspberry Pi, they're both exposed boards with ARM processors and are HD video competent. In any case, BeagleBoards and PandaBoards have more connectors and association headers (pieces of the board that can be utilized by fastening extra equipment) than the Raspberry Pi, and the two gadgets are somewhat bigger. Coming up next aren't thorough arrangements of segments, yet a few highlights contrast from the Pi. For its planned instructive purposes, the Raspberry Pi has two significant benefits over the others. To start with, it is considered to be a finished working PC. You basically need to embed a SD card containing the OS, interface the peripherals and force, and it's all set. BeagleBoards and PandaBoards expect hookup to a host PC for introductory arrangement. In spite of the fact that they have comparable handling abilities, they take somewhat more expertise to get them completely utilitarian.

Table 1. Comparison between BeagleBoard and PandaBoard

| BeagleBoardandBeagleBoard-xM | PandaBoard and PandaBoard ES |
|---|---|
| Cortex A8-based processor madeby Texas Instruments, running from 600 MHz to 720 MHz on the BeagleBoard (depending upon version) and 1 GHz on the xM. | Dual-core ARM Cortex A9 MPCore processor, also manufactured by Texas Instruments, running at 1GHz on the PandaBoard and 1.2 GHz on the ES |
| 128 MB RAM on the original BeagleBoard, but 256 MB and 512 MB RAM on the newer boards, respectively. | 1 GB RAM |
| DVI-D monitor connector<br>S-video connector<br>Audio in and out (not just audio out)<br>One USB port on BeagleBoard and four USB ports on the xM<br>USB and DC power<br>No RCA or HDMI connector | DVI-D monitor connector<br>LCD expansion header<br>Audio in and out<br>One USB on-the-go port and two standard USB ports<br>WiFi and Bluetooth connectivity<br>USB and DC power<br>No RCA connector |

Second, the other devices are much more expensive than the Raspberry Pi. For example, in April 2012, the pricing was $125 to $149 for the two main BeagleBoard models, and $174 to $182 each for the two PandaBoard models. These prices are a far cry from the $25 and $35 base prices of the Raspberry Pi. Given its functionality and price, the Raspberry Pi seemed better poised for us.

### B. Smart Speaker Architecture

KASPER provides access to KASPER on Linux distributions on desktop as well as hardware devices like Raspberry Pi. It is a headless client that can be used to interact with KASPER via voice only. As more and more features like multiple hotword detection support and wake button support were added to KASPER Linux, the code became complex to understand and manage. A system was needed to model the app after. The Finite State Machine was the perfect approach for such a system.

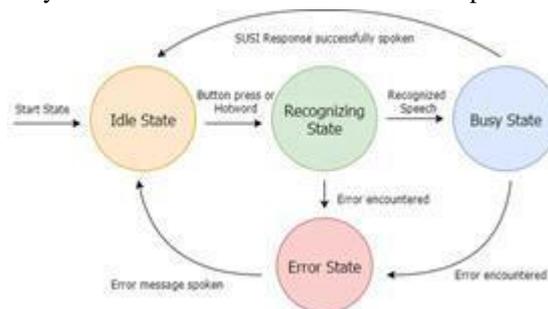

Fig 2. The finite State Machine

The Wikipedia meaning of a State Machine is "It is a theoretical machine that can be in precisely one of a finite

355

number of states at some random time. The FSM can change starting with one state then onto the next because of some outside information sources; the change starting with one state then onto the next is known as a progress."

This implies that in the event that you can demonstrate your application into a finite number of states, you may consider utilizing the State Machine usage.

State Machine execution has the accompanying focal points:
Better authority over the working of the application.
Improved Error taking care of by making an Error State to deal with mistakes.
States work freely which serves to modularize code in a superior structure.

Regardless, we announce a theoretical State class. This class proclaims all the basic properties of a state and change technique.

We declared the on_enter() and on_exit() conceptual method. These methods are executed on entering and exiting a state separately. The errand assigned for the state can be acted in the on_enter() method and it can let loose assets or quit tuning in to callbacks in the on_exit() method. The transition method is to transition starting with one state then onto the next. In a state machine, a state can transition to one of the permitted states only. Hence, we check if the transition is permitted or not prior to continuing it. The on_enter() and transition() methods additionally acknowledge a payload contention. This can be utilized to move some information to the state from the past state. We likewise added the components property to the State. Components store the common components that can be utilized across all the State and should have been initialized only once. We make a component class pronouncing all the components that should have been utilized by states.

● **Idle State:** App is listening for Hotword or Wake Button.
● **Recognizing State:** App actively records audio from Microphone and performs Speech Recognition.
● **Busy State:** KASPER API is called for the response of the query and the reply is spoken.
● **Error State:** Upon any error in the above state, control transfers to Error State. This state needs to handle the speak the correct error message and then move the machine to Idle State.

Each state can be implemented by inheriting the base State class and implementing the on_enter() and on_exit() methods to implement the correct behavior.

We also declare a KASPER State Machine class to store the information about current state and declare the valid transitions for all the states.

We also set Idle State as the current State of the System. In this way, the State Machine approach is implemented in KASPER Linux.

**C. Modified FSM Architecture**

During the underlying phases of KASPER: As the code base developed, it was getting hard to maintain code, so we selected to execute a Finite State Architecture in our repo. However, as there were new highlights actualised in the codebase, we understood that we were unable to deal with more than each inquiry in turn which limited a great deal of highlights. eg. The smart speaker was changed over to a basic Bluetooth speaker since no reaction with respect to playing/stopping were acknowledged.

To settle this issue, we made a slight alteration in the architecture.

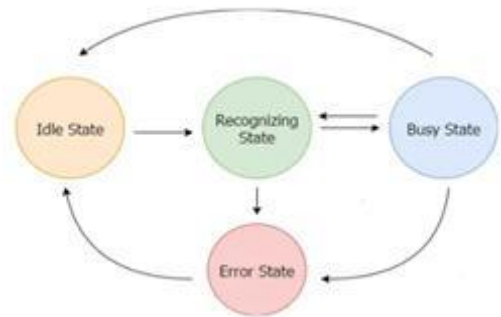

Figure 3. The Modified finite State Machine

*1) Adding a Second Hotword Recognition Class*

To enable KASPER to handle the simultaneous question, The State machine should be set off while KASPER is giving out the principal reaction and to trigger the State Machine, we should have hotword recognition while KASPER is talking the response to the past inquiry. Thus, a hotword recognition motor is presently started each time the State Machine enters the bustling state.

*2) Modifying the State Machine Architecture*

After declaring pronouncing a second hotword recognition engine , we altered how the changes occur between the States of the KASPER State Machine.

Subsequently, the callback that was set off was passed from the busy state.

When the hotword is distinguished ,the state machine makes advances to the Recognition State while stopping the current Music and resumes the Music after the subsequent question has been finished.

This is the manner by which KASPER measures different questions at the same time while still keeping up finite state architecture.

**D. Speech to text recognition**

Speech Recognition is basically making the computer understand what we speak. In general, a computer such that it can hear us and respond back to us and by "understand" we mean it would convert the speech into appropriate text. Thus speech recognition is also called the Speech to Text conversion process. It consists of a microphone for humans to speak, recognition of speech software, and a computer to

356

perform tasks. The basic recognition of the speech system is shown.

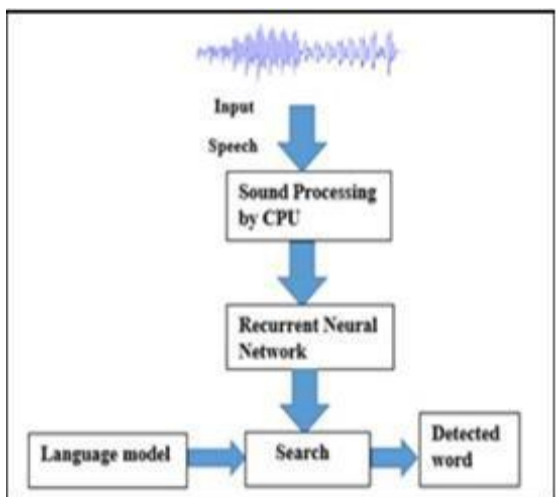

Figure 4. Speech Processing model of Kasper AI

*1) Speech to text engine*

Speech to text engines is used to feed sound waves into the computer for converting them into text. As sound waves are continuous (analog) signals the first thing is to do a sampling of the signal using the Nyquist theorem. This sampled signal is fed directly to our neural network but pre-processing of the signal is done in order to get better results and accurate predictions of spoken words. In pre-processing, we grouped large sampled signals into 20-millisecond small chunks. Preprocessed data which is in digital format is now fed to our Recurrent Neural Network (RNN) which is our main speech recognition model and it is based on many to many architecture which is used for prediction.

*2) Sampling and pre-processing of speech*

Sampling and pre-processing of data is an important step while designing STT Engine. This step decides the performance and time consumed by the engine. Sound waves are as we know one-dimensional. At every point in time, they have a single value based on the magnitude of the wave. To turn this sound wave into numbers just record the magnitude of the wave at equally-spaced points. This is called Sampling. This sampled data is directly fed into our recurrent neural network but for ease and better results data is preprocessed before applying to the network. Pre-processing is breaking the sampled data into smaller groups of data. Generally grouping the wave within some interval of time mostly for 20-25 milliseconds. Sampling and preprocessing together can be termed as the conversion of sound into numbers(bits).

*3) Recurrent Neural Network*

Now audio is given at input and is easy to process, it will be fed to our deep neural network. After feeding these small audio chunks of 20ms to our network it will figure out a letter that matches the spoken sound. RNN is a network that has a memory that decides future predictions. This is because as it predicts a letter it will affect the likelihood of the next letter which it will predict. Consider an example, if we have said "MUM" so far, then it obviously will predict "BAI" next to complete the word "MUMBAI". There is less probability that one will say something which is completely out of context such as "ABC" after saying the word "MUM". Hence having a memory of previous predictions boosts our network to make more accurate predictions going forward. RNNs use the idea of sequential information. RNN is a neural network that has a memory that influences future predictions sequential information which is stored into the memory of RNNs and is used for predictions. The idea to use RNN instead of a traditional neural network is in traditional neural networks, it is assumed that every input & output doesn't depend on each other. Hence using a traditional neural network is a bad idea in speech processing. Prediction of any word in a sentence requires the information about the word which came before i.e. past word which is processed. Having a memory is one of the specialties of RNN that makes it unique from other networks. There are various neural networks available among them the Recurrent Neural network is used because it is more efficient than the others for speech recognition.

*4) Various Stt Engines Using RNN*

There are various engines used that are based on RNN"s which uses python, C, java programming languages to build a Recurrent Neural Network. We have gone through CMU pocket-sphinx, snowboy hot word detection which uses python language, and RNN which shows good results also if we increase the database it will perform at it"s best. Unlike Google"s STT and Amazon"s Alexa, CMU pocket-sphinx is an offline speech to text conversion engine provided that training of a dataset is done online. CMU online training portal must be given a set of words that we have to train. The training process is the same as discussed above in training at RNN. It uses python to build an LSTM network. Also recently launched engine Snowboy-hot word detection works offline but it is limited to the detection of one particular hot word.

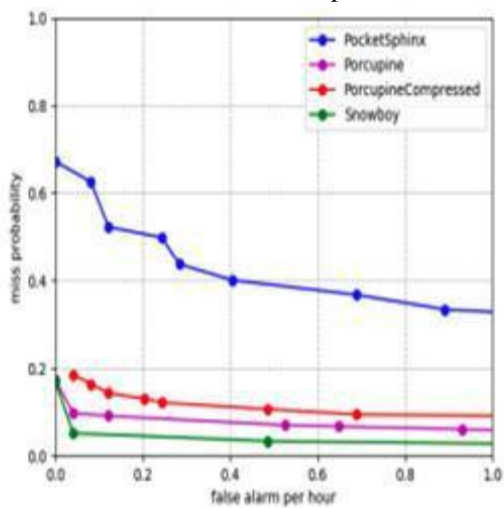

Fig 5. Comparison of Wake Word Engines

## 3. Extracting Context

After the Speech to text engine converts the audio input into text, the system needs to make some sense out of that. The aim is to find the intent of the audio query using some algorithm. A convolutional neural network (CNN, or ConvNet) falls under a class of deep neural networks, which is most commonly used to analyze visual imagery. Some common applications of CNN are image classification , facial recognition , object detection etc. Most recently, however, CNN have also found to perform well with problems associated with NLP tasks such as Sentence Classification, Sentiment Analysis, Text Classification, Text Summarization, Machine Translation and Answer Relations. Hence we decided to use CNN.

A. General architecture of a Convolutional Neural Network

A CNN is composed of "convolutional" layers and "downsampling" or "subsampling" layers.

Pooling layers or downsampling layers are often placed after convolutional layers in a ConvNet, mainly to reduce the feature map dimensionality for efficiency, which in turn improves the actual performance. Convolutional layers consist of neurons that scan their input for patterns.

Generally, the two layers, i.e., Pooling and Convolutional layers, are in an alternate order, but that's not a necessary condition.

This is followed by a Multi Layer Perceptron with one or more layers.

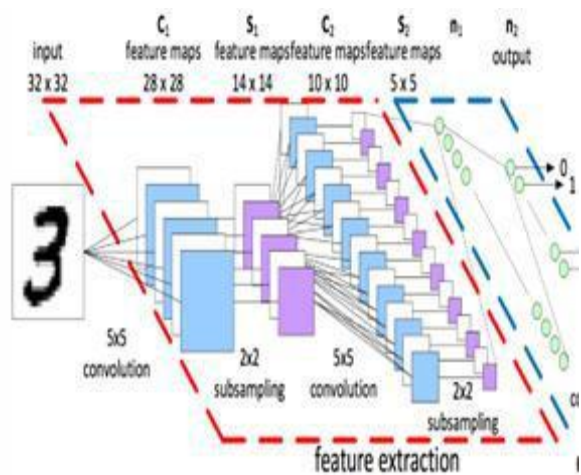

Fig 6. CNN Architecture

We compared a variety of algorithms to get the best results possible:
1. Dynamic Programming based Fuzzy Approach
2. K- Nearest Neighbors Algorithm
3. Convolution Neural Networks
4. Recurrent Neural Networks

*1) Fuzzy Matching:*
Dynamic Programming is a mathematical procedure of optimization using multistage decision progression. It is a general algorithm design method for solving problems formulated as repetitions with overlapping sub instances. In 1965 fuzzy set theory was developed by Lotfi A. Zadeh, which has become a significant device in dealing with roughness and imprecision in real-world problems. In 1970 Bellman applied fuzzy set theory in decision-making problems. The technique of solving optimization problems in dynamic programming involving fuzzy parameters is known as fuzzy dynamic programming. In our case, we used fuzzy DP and got an accuracy of 43.29.

*2) KNN(K Nearest Neighbors):*
K-nearest neighbors (KNN) algorithm comes in the category of supervised ML algorithm used for both classifications and predictive regression problems. However, it is primarily used for classification predictive problems in the industry. KNN uses feature similarity to predict new data points values, which means that the new data points will be assigned a value based on how closely it matches the training set's points. In our case, we used KNN to define a minimum accuracy above which all the other algorithms perform. The accuracy we got with KNN was 51.37.

*3) CNN(Convolutional Neural Networks):*
CNN is a deep, feed-forward artificial neural network where connections between nodes do not form a cycle. CNN's are generally used in computer vision; however, they've shown promising results when applied to various NLP tasks as well. CNN's are good at extracting local and position-invariant features, whereas RNN's are better when classification is determined by a long-range semantic dependency rather than some local key-phrases. For tasks where feature detection in the text is more important, for example, searching for angry terms, sadness, abuses, named entities, etc. CNN's work well, whereas for tasks where sequential modeling is more critical, RNNs work better.

In our case, CNN had an accuracy of 78.23%.

*4) RNN(Recurrent Neural Networks):*
RNN comes under Neural Network category where the output from the last step is fed back as an input to the current step. It is a sequence of neural network blocks that are linked to each other like a chain. This allows RNN to exhibit temporal behavior and capture sequential data and is trained to recognize patterns across time, making it a more 'natural' approach when dealing with textual data since the text is naturally sequential. However, in our case, there wasn't any significant accuracy difference between RNN and CNN because text classification doesn't need to use the information stored in the sequential nature of the data. A big argument for CNNs is that they are much faster (~5x) than RNNs in computation time.

In our case, RNN had an accuracy of 72.71%.

**How does the CNN architecture work for Sentence Classification?**

Just like images can be represented as an array of pixel values. Similarly, text can be represented as an array of vectors where each word is mapped to a specific vector in a vector space composed of the entire vocabulary that can be processed with the help of a CNN. When we are working with sequential data, like text, we work with one-dimensional convolutions, though the idea and the application stay the same. We still want to pick up patterns in the sequence that become more complex with each added convolutional layer. Here, we are going to train a Convolutional Neural Network to perform sentence classification on data that we get after converting speech to text.

We followed the following workflow:
- We import the data and preprocess it into a desirable format(one we can work with) using pandas.
- We use GloVe to obtain pre-trained word embeddings for our model.
- Keras is used to train the data on a CNN architecture and evaluate the accuracy obtained on the validation set.

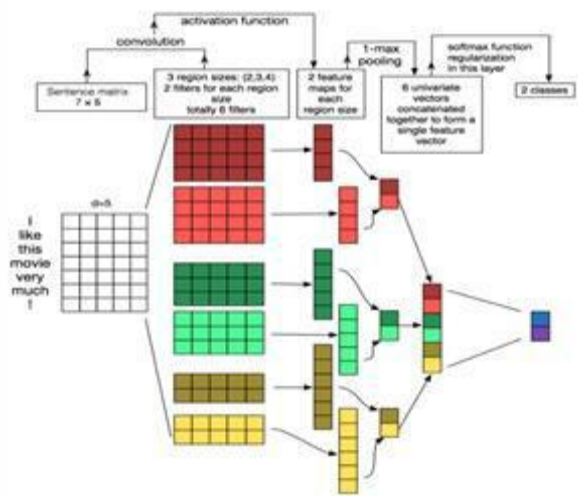

Fig 7. Context Inference

C. Our Classes

Using CNN, we were able to classify sentences into 22 different classes, which are:
- Art and Beauty
- Business and Finance
- Communication
- Connected Car
- Food and Drink
- Games, Trivia, and Accessories
- Health and Fitness
- Interests
- Knowledge
- Lifestyle
- Movies and TV Shows
- Music and Audio
- News
- Novelty and Humour
- Problem Solving
- Productivity
- Shopping
- Social
- Sports
- Travel and Transportation
- Utilities
- Weather

4. Result

A privacy enabled smart speaker prototype was created. The prototype is "smart", makes coherent conversations and can be customized according to the user's requirements. A privacy focussed smart speaker was created, giving the user the ability to control the server and the client code.

5. Conclusion

Over 50% of the Indian population are expected to own a regular smart speaker by 2023, and it's predicted that smart speaker ownership would overtake tablets globally by 2022. This paper provides an overview of the technology and tools required to build a privacy first custom smart speaker.

Kasper.AI is able to provide a custom skills system trained with CNNs to allow the speaker to be "smart" and make coherent conversations